\documentclass[traditabstract]{aa} 
\usepackage{graphicx}
\usepackage{txfonts}
 \usepackage[T1]{fontenc}
\usepackage{color}
\usepackage[]{natbib}
\usepackage[switch]{lineno}

        \newcommand{\msun}{\ensuremath{\mathrm{M}_{\odot}}}
        
        \newcommand{\sfr}{M$_{\odot}$ yr$^{-1}$}
        
        \newcommand{\ha}{${\rm H\alpha}$}
        \newcommand{\han}{${\rm H\alpha_{\rm nar}}$}
        \newcommand{\oiii}{[O{\sc iii}]$\lambda$5007}
        \newcommand{\nii}{[N{\sc ii}]}
        \newcommand{\niin}{[N{\sc ii}]$_{nar}$}

        \newcommand{\oiiin}{[O{\sc iii}]$_{nar}$}
        \newcommand{\hb}{${\rm H\beta}$}
        
        \newcommand{\lagn}{${\rm L_{AGN} }$}

        \newcommand{\pap}{Paper I}

        \newcommand{\mstar}{M$_{\star}$}

\begin{document}

   \title{Fast outflows and star formation quenching in quasar host galaxies\thanks{based on observations collected  at the European Organisation for Astronomical Research in the Southern Hemisphere, Chile, P.ID: 086.B-0579(A) and  091.A-0261(A)  }}

   \author{S. Carniani \inst{1,2,3,4},
          A. Marconi  \inst{1,2},
          R. Maiolino \inst{3,4},
          B. Balmaverde \inst{1},
          M. Brusa \inst{5,6},
          M. Cano-D\'iaz \inst{7}
          C. Cicone   \inst{8},
          A.~Comastri \inst{6},
          G. Cresci  \inst{2},
          F. Fiore \inst{9},
          C. Feruglio \inst{9,10,11},
          F. La Franca \inst{12},
          V. Mainieri \inst{13},
          F. Mannucci \inst{2},
          T. Nagao \inst{14},
          H.~Netzer \inst{15},
          E. Piconcelli \inst{9},
          G. Risaliti \inst{2},
          R. Schneider \inst{9},
          O. Shemmer \inst{16}
                 }

   \institute{Dipartimento di Fisica e Astronomia, Universit\`a di Firenze, Via G. Sansone 1, I-50019, Sesto Fiorentino (Firenze), Italy 
                      \and
             INAF - Osservatorio Astrofisico di Arcetri, Largo E. Fermi 5, I-50125, Firenze, Italy 
                      \and
        Cavendish Laboratory, University of Cambridge, 19 J. J. Thomson Ave., Cambridge CB3 0HE, UK 
                \and
        Kavli Institute for Cosmology, University of Cambridge, Madingley Road, Cambridge CB3 0HA, UK 
                \and
        Dipartimento di Fisica e Astronomia, Universit\`a di Bologna, viale Berti Pichat 6/2, 40127 Bologna, Italy  
        \and 
        INAF - Osservatorio Astronomico di Bologna, via Ranzani 1, 40127 Bologna, Italy 
        \and
        Instituto de Astronom\'{\i}a, Universidad Nacional Aut\'onoma de M\'exico, Apartado Postal 70-264, Mexico D.F., 04510 Mexico 
        \and    
         Institute for Astronomy, Department of Physics, ETH Zurich, Wolfgang-Pauli-Strasse 27, CH-8093 Zurich, Switzerland 
                \and
        INAF - Osservatorio Astronomico di Roma, via Frascati 33, 00040 Monteporzio Catone, Italy 
        \and
        Scuola Normale Superiore, Piazza dei Cavalieri 7, 56126 Pisa, Italy 
        \and 
 IRAM - Institut de RadioAstronomie Millime\'trique, 300 rue de la Piscine, 38406 Saint Martin d'He\'res, France 
         \and
         Dipartimento di Matematica e Fisica, Universit\'a Roma Tre, via della Vasca Navale 84, I-00146 Roma, Italy 
        \and 
        European Southern Observatory, Karl-Schwarzschild-str. 2, 85748 Garching bei München, Germany 
        \and
        Research Center for Space and Cosmic Evolution, Ehime University, Bunkyo-cho 2-5, Matsuyama, 790-8577 Ehime, Japan 
        \and
        School of Physics and Astronomy, The Sackler Faculty of Exact Sciences, Tel-Aviv University, Tel-Aviv 69978, Israel 
        \and
        Department of Physics, University of North Texas, Denton, TX 76203, USA; ohad@unt.edu 
                  }

   \date{Received; accepted}

\abstract{Negative feedback from active galactic nuclei (AGN) is considered a key  mechanism in shaping galaxy evolution.
Fast, extended outflows are frequently detected in the AGN host galaxies at all redshifts and luminosities, both in ionised and molecular gas.
However, these outflows are only potentially able to quench star formation, and we are still lacking decisive evidence of negative feedback in action.
Here we present observations obtained with the Spectrograph for INtegral Field Observations in the Near Infrared (SINFONI) H- and K-band integral-field of two quasars at $z\sim$2.4 that are
characterised by fast, extended outflows  detected through the \oiii\ line. The high signal-to-noise ratio of our observations allows us to identify faint narrow (FWHM $< 500$ km/s) and spatially extended components in \oiii\ and \ha\ emission  associated with star formation in the host galaxy.
This star formation powered emission is spatially anti-correlated with the fast outflows. 
The ionised outflows therefore appear to be able to suppress star formation in the region where the outflow is expanding. However, the detection of narrow spatially extended \ha\ emission indicates  star formation rates of at least $\sim 50-90$ \sfr, suggesting either that AGN feedback does not affect the whole galaxy or that many feedback episodes are required before star formation is completely quenched.
On the other hand, the narrow \ha\ emission extending along the  edges of the outflow cone may also lead  also to a positive feedback interpretation.
Our results highlight the possible double role of galaxy-wide outflows in host galaxy evolution. 
}

\authorrunning{Carniani et al.}
\titlerunning{Fast outflows quenching star formation in quasar host galaxies}
 \maketitle

\section{Introduction}
A key  problem in galaxy formation and evolution is understanding how active galactic nuclei (AGN) interact with their host galaxies. In particular, the origin of the tight  correlations observed between the masses of supermassive black holes (BHs) and the stellar masses, stellar velocity dispersions, and luminosities of the host galaxy bulges is still debated (e.g. \citealt{Magorrian:1998, ferrarese:2000, gebhardt:2000, Marconi:2003, Kormendy:2013, Reines:2015}). 
Negative feedback from AGN can provide a viable explanation for these correlations \citep{Fabian:2012, King:2015}.
According to theoretical models, AGN-driven outflows regulate  BH growth and  star formation activity in the host galaxies by blowing away the  gas that feeds star formation and BH growth. Not only does negative feedback  provide the link between the growth of supermassive black holes and their host galaxies, it also is a fundamental mechanism to explain the steep slope of the high end of the stellar mass function and to account for the existence of the red sequence of massive passive galaxies (e.g. \citealt{Baldry:2004, Hopkins:2006, Perez-Gonzalez:2008}).

Several theoretical models have been proposed for radiation-pressure-driven outflows in AGN as possible feedback mechanism able to regulate BH and stellar mass accretion in active galaxies \citep[e.g.][]{Granato:2004,Di-Matteo:2005,Menci:2008, King:2010, Zubovas:2012,Fabian:2012, Faucher-Giguere:2012, Zubovas:2014,Nayakshin:2014, Costa:2014, Costa:2015, King:2015}.
Although AGN-driven outflows should be powerful enough to sweep away all gas from the host galaxy, recent simulations \citep[e.g.][]{Roos:2015} and observations \citep[e.g.][]{Balmaverde:2016} indicate that the effect of AGN feedback on the star formation in the host galaxy might be marginal. The effect of the feedback mechanism on the star formation history and host galaxy evolution is still an open question.

There is much evidence of AGN outflows of both molecular and ionised gas \citep[e.g.][]{Cicone:2012, Maiolino:2012,  Cano-Diaz:2012, Rupke:2013, Feruglio:2013, Feruglio:2013a, Cicone:2014,Harrison:2014, Aalto:2015, Feruglio:2015, Brusa:2015, Perna:2015, Perna:2015a, Cresci:2015, Cresci:2015a, Cicone:2015, Carniani:2015, Harrison:2016}. 
 Molecular and ionised outflows are extended from a few hundred pc \citep[e.g.][]{Feruglio:2015} to tens of kpc \citep[e.g.][]{Harrison:2012,Cresci:2015} and are characterised by mass-loss rates that can exceed the star formation rate (SFR) by two orders of magnitude \citep{Cicone:2014}.
In particular, when molecular outflow rates are compared with molecular gas masses, gas depletion timescales (i.e. the time required for the outflow to completely expel the available gas) are much shorter than typical galaxy timescales, indicating that outflows are \textit{\textup{potentially}}
able to provide the negative feedback required by the models \citep[e.g.][]{Cicone:2014}. 
This means that while these results suggest that molecular AGN-driven outflows can potentially quench star formation,  we are still lacking direct evidence that they are effectively doing so.

As yet, there have only been a few tentative detections of the signature of AGN exerting a negative feedback on star formation in the host galaxy. In a luminous QSO at  $z\sim2.5$, \cite{Cano-Diaz:2012} presented evidence that \ha\ emission tracing star formation is suppressed in the region affected by fast outflows that is
traced by \oiii\ emission. A similar result  was found by \cite{Cresci:2015} in an obscured QSO at $z\sim1.5$, where the ionised outflow appears to be sweeping away the gas in the host galaxy and thus quenches star formation.  
Given the low signal-to-noise ratio (S/N) of the star-formation-powered \ha\ detected by \cite{Cano-Diaz:2012} and by \cite{Cresci:2015}, we need to confirm these findings with higher S/N and in a larger sample of objects.
Moreover, the \ha\ emission in the two host galaxies of these QSOs indicates remarkably high SFRs of $\sim100$ \sfr.
Therefore, it is still unclear whether these AGN-driven outflows are able to \textit{\textup{completely}} quench SF in the host galaxy, as required by models.

In this work we present new observations made with the Spectrograph for INtegral Field Observations in the Near Infrared (SINFONI) K-band of two quasars at $z\sim2.4$: LBQS 0109+0213 (hereafter LBQS0109) and HB89 0329-385 (hereafter HB8903).  These objects are part of a larger sample observed with SINFONI  in the H band with the aim of mapping the kinematics of the \oiii\ line (Carniani et al. 2015, hereafter Paper I).
Their spatially resolved \oiii\ kinematics clearly reveals extended ionised outflows with velocities  exceeding $1000$ km/s. In addition to the broad and bright outflowing component of \oiii, these objects are characterised by faint, narrow \oiii\ lines that
are most likely associated with star formation powered emission in the host galaxies.
The aim of the Very Large Telescope (VLT)/SINFONI K-band observations is to detect the corresponding narrow \ha\ components to verify the star formation origin and study its spatial distribution compared to  the outflow location, as was done in \cite{Cano-Diaz:2012} and \cite{Cresci:2015}.

The paper is organised as follows: in Sect. 2 we present the properties of the K SINFONI datasets, and in Sect. 3 we show the analysis of the H- and K-band spectra aimed at revealing narrow \oiii\ and \ha\ components. The spatial distribution of the two narrow components in both QSOs is discussed in Sect. 4. Finally, the conclusions are presented in Sect. 5.  A $H_0 = 67.3$ km s$^{-1}$ Mpc$^{-1}$, $\Omega_{M} = 0.315$,  $\Omega_{\Lambda} = 0.685$ cosmology is adopted throughout this work \citep{Planck-Collaboration:2014aa}.

\section{Observations and data reduction}\label{sec:data}
The two sources LBQS0109 and HB8903 were observed with the SINFONI at the VLT. Observations from program 091.A-0261(A) (PI A. Marconi) were executed  in seeing-limited mode ($\le 0\farcs6$) with 0\farcs250 spatial scale and the K-band grating, with a spectral resolution of  R = 4000  over the observed wavelength range $\lambda=1.95-2.45$ nm.
The observations of each target were divided into six 1h long observing blocks (OB) for a total on-source integration time of four hours (plus overheads). During each observing block, an ABBA nodding was adopted to perform  sky subtraction. Standard stars for telluric correction and flux calibration were observed shortly after or before the execution of each OB.  
K-band data were reduced  using the standard ESO-SINFONI pipeline  2.6.0 after removing cosmic-ray hits from the raw data with the L.A. Cosmic procedure by \cite{van-Dokkum:2001}. Sky subtraction was optimised using the IDL routine by \cite{Davies:2007}. The final data cubes have a spatial scale of 0\farcs125$\times$0\farcs125 and a field of view of 8\arcsec$\times$8\arcsec. 
However, all figures in this paper will only show a fraction of the field of view (2\arcsec$\times$2\arcsec), since  observed  emissions are extended for less than $\sim$1\arcsec\ ($\sim$8 kpc).
The estimated angular resolution is $\sim0\farcs6$, based on a 2D Gaussian fitting of the flux map of the spatially unresolved broad \ha\ line (see Sect. 4 for a description of the fitting procedure).

In addition to the K-band dataset, we also considered the H-band SINFONI spectra from  program 086.B-0579(A), which are described in  detail in  \pap. 

\begin{table*}
\caption{Quasar properties.}           
\label{tab:summary}      
\centering          
\begin{tabular}{l c c c c c c  c  c c }    
                         \\
\hline
QSO &   $z$ & RA & DEC  & Log$_{10}$($\frac{L_{\rm AGN}}{\rm erg/s})$ & $v_{\rm outflow}$   & $R_{\rm outflow}$  &$M^{outflow}_{\rm [OIII]}$      & $  \dot M^{outflow}_{\rm [OIII]}$ & $M_{\rm BH}$  \\ 
   &  &  & & & [km/s] & [kpc] & [$10^7$ \msun] & [\sfr ] & [$10^{10}$ \msun] \\
 (1)  & (2) & (3) & (4) & (5) & (6) & (7) & (8) & (9) & (10)\\

\hline
\\
LBQS0109  & 2.35  & 01:12:16.9 & +02:29:47   &  47.5  & 1850 & 0.4 & 1.2 & 60 & 1.0    \\
\\
HB8903 & 2.44  &  03:31:06.41 & -38:24:04.6   & 47.5  & 1450 & 1.9 & 0.7 & 6 & 1.3 \\
\\
\hline

\\
                
\end{tabular}   
\tablefoot{  (1) ID of the object (2) Red-shifts are estimated from \oiii. (3,4) Coordinates (5) AGN bolometric luminosities are derived  using the relation  \lagn \ $\sim 6\,\lambda L(\lambda5100\AA)$  from \cite{Marconi:2004}. (6, 7) Outflow velocities and radii are inferred with the spectroastrometric method described in \pap. (8)  Outflow masses are inferred from \oiii\ assuming a ${\rm T_e \sim 10^4}$ K and a ${\rm n_e \sim 500 cm^{-3}}$. (9) Outflow mass rates are calculated as $\dot M^{outflow} = M^{outflow} v_{outflow}\,/\,R_{outflow}$. (10) Black hole masses are adopted
from \cite{Shemmer:2004}. }

\end{table*}

\section{Data analysis}

In \pap\ we presented H-band SINFONI spectra of  five QSOs at $z\sim2.4$ that were used to map extended outflows traced by the \oiii\ emission line. The broad blue wings of the \oiii\ line indicated outflow velocities  of 500-2300 km/s. 
By applying a new method, based on  spectroastrometry of the \oiii\ line, we found that the spatial centroids of the emission extracted from blue-shifted spectral channels are displaced with  respect to  the QSO position,  suggesting that the outflowing gas is extended a few kpc from the centre and is not isotropically distributed (Fig. 4 of \pap). We inferred ionised outflow rates of $\sim 10-700$ \sfr.  Both outflow velocities and mass rates increase with AGN bolometric luminosity, suggesting a  correlation between the two quantities. In this work we present new  observations of two QSOs, LBQS0109 and HB8903, belonging to the sample of \pap,\ whose characteristics are summarised in Table~\ref{tab:summary}.

\subsection{Narrow \oiii\ components}\label{sec:oiii}

  \begin{figure*}
   \centering 
\includegraphics[width =1\textwidth]{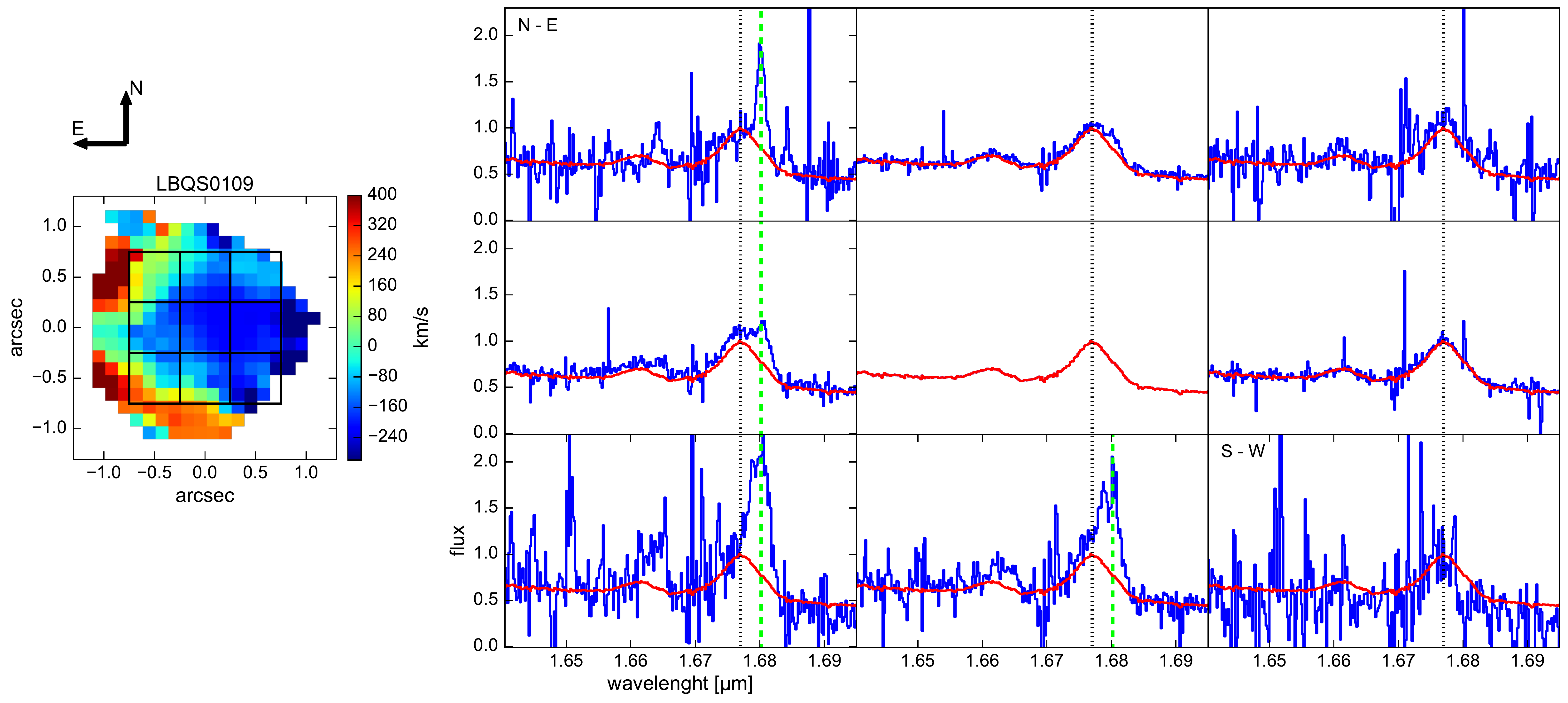}
\includegraphics[width =1\textwidth]{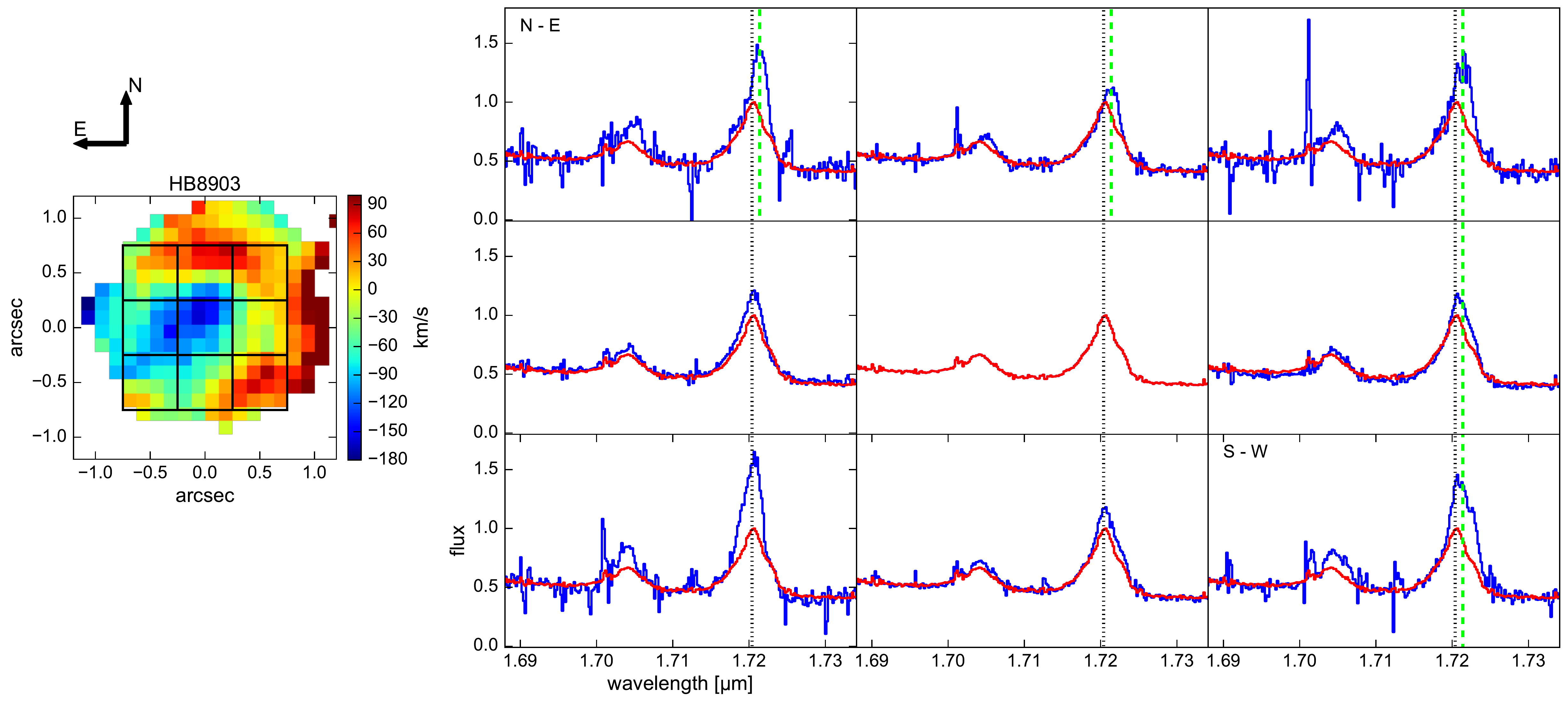}

\caption{ 
\emph{Left panels}: the \oiii\ velocity maps of LBQS0109 (top) and HB8903 (bottom) presented in \pap. Black squares indicate the nine regions where we extracted the corresponding spectra shown in the right panels. 
\emph{Right panels}: comparison between the nuclear \oiii\ profiles (red) and those from the external regions (blue) of LBQS0109 (top) and HB8903 (bottom). 
The black dotted line corresponds to the \oiii\ central wavelength.
Narrow (FWHM$\sim$490 km/s for LBQS0109 and FWHM$\sim$480 km/s for HB8903) components of the \oiii\ emission, identified by the green dashed line,  are clearly detected. }

 \label{fig:narrow_oiii}
   \end{figure*}

\begin{figure}
   \centering
\includegraphics[width =0.45\textwidth]{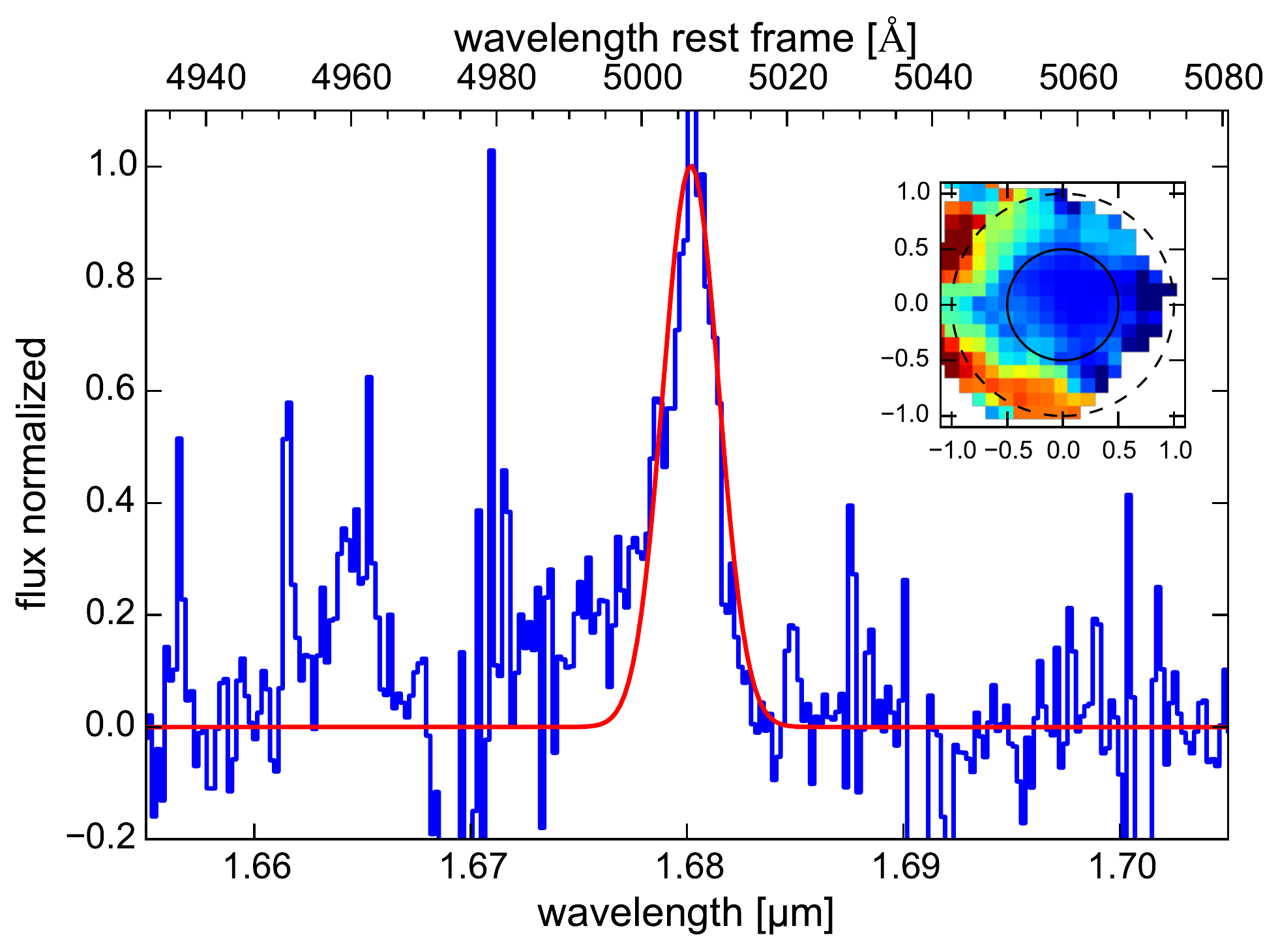}\\
\includegraphics[width =0.45\textwidth]{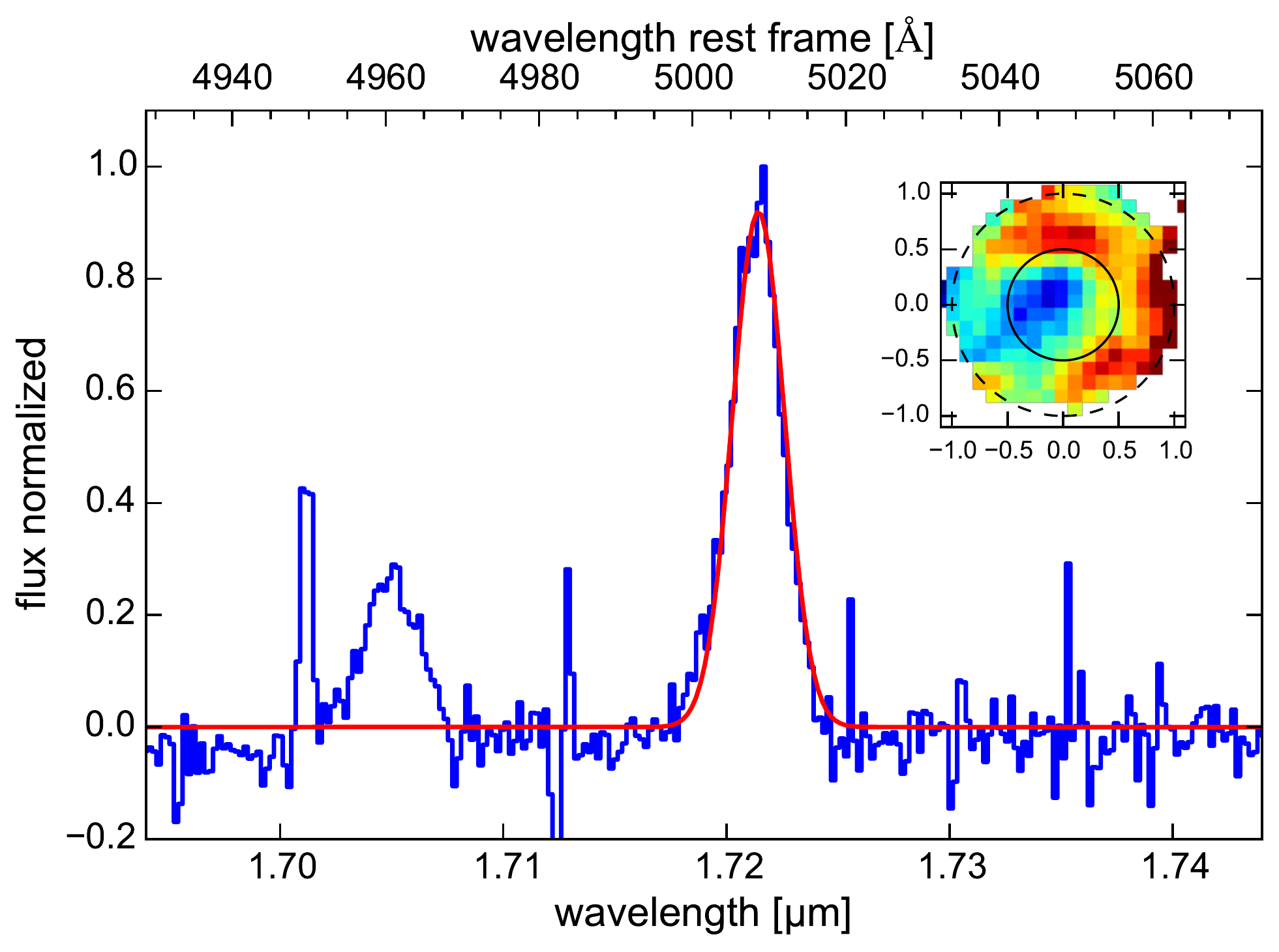}
\caption{ H-band SINFONI spectra of LBQS0109 (top) and HB8903 (bottom).
The observed spectra (blue lines) are the result of the subtraction between the spectra extracted from the ring-shaped regions (0.5\arcsec$< r <$ 1\arcsec )  and from the nuclear regions (r $<$ 0.5\arcsec). The narrow doublet \oiii\ is evident in both sources. The red curves denote the Gaussian fits whose parameters are shown in Table \ref{tab:results}. In the smaller inset, the nuclear region (solid black line) and the ring-shaped region (dashed black line) are drawn in the \oiii\ velocity maps.}
 \label{fig:spectra_oiii}
   \end{figure}

In the two QSOs, we  detect a weak, narrow (FWHM $<$ 500 km/s) component of the \oiii\ emission that is only visible  in some parts of the field of view.  The narrow components are on the red side of the \oiii\ profile.   In Fig. \ref{fig:narrow_oiii}  we compare for both QSO host galaxies the spectrum  extracted from a region of $\sim$4kpc~$\times$~4kpc centred on the AGN position (red profiles) with the spectra extracted from external regions of the host galaxy (blue profiles).  The regions from which we extracted the spectra are shown in the left panels of Fig.  \ref{fig:narrow_oiii}.
The narrow \oiii\ component (hereafter \oiiin), identified in the figure by a dashed green line, is clearly visible in the  eastern (E) side for LBQS0109 and in the  western (W) side for HB8903, showing that in these regions the \oiii\ profile is different from the one extracted from the nuclear region.  
We note that the blue-shifted narrow emission detected in the lower left boxes of HB8903 is mostly due to outflow emission since the velocity dispersion in this region (FWHM $\sim$ 800 km/s) is larger than that observed in the other external regions (FWHM $\sim$ 500 km/s); we refer to \pap\ for more details.
In \pap\ we showed that a broken power-law parametric profile represents a good fit to the main \oiii\ component
and, in particular, that it is sufficient to reproduce the spectra of nuclear and outflow regions (e.g. Fig.~1 of \pap). 
However, a more complex line profile should be used  to describe the spectra extracted from  external regions where  \oiiin\ is visible (Fig.~\ref{fig:narrow_oiii}).
For this reason, we reanalysed the H-band spectra and performed a new pixel-by-pixel fitting by adding an additional narrow, red-shifted component to the parametric profile used in \pap.
Since the S/N of  \oiiin\ in each spatial pixel is low (S/N$<9$), we cannot perform a kinematical analysis of this new component. Therefore, its velocity and width were fixed to the values obtained from a Gaussian fit of the line profile obtained by subtracting the nuclear spectrum ($ r <0.5\arcsec$ ) from the circumnuclear spectrum ($0.5\arcsec< r <1\arcsec $),
after re-scaling  to match the intensities of the broad \hb\ and \oiii\ lines 
(Fig. \ref{fig:spectra_oiii}) .
The best-fit  results are listed in Table \ref{tab:results}  for both targets. 

\begin{table*}
\caption{Central wavelengths and line widths of the two narrow components of \oiiin\ and \han.}      
\label{tab:results}      
\centering          
\begin{tabular}{ c   c c | c c }    
                         \\ 
\hline
 & \multicolumn{2}{c}{\oiiin} &  \multicolumn{2}{c}{\han} \\
 \hline
 & $\lambda_0$  &  FWHM &  $\lambda_0$  &  FWHM \\
  & ${\rm [\mu m]}$  &  [km/s] & ${\rm [\mu m]}$  &  [km/s]  \\
   \hline

LBQS0109 & 1.6802$\pm$0.0004 & 490$\pm$90 &  2.203$\pm$0.001 &   250$\pm$200 \\
HB8903 & 1.7214$\pm$0.0002 & 480$\pm$70 &  2.2560$\pm$0.0006 & 500$\pm$150 \\

\hline  
\end{tabular}
        
\end{table*}

The flux maps of the \oiiin\ components are shown in Fig.~\ref{fig:map_oiii}. The narrow emission is not symmetrically distributed around the QSO. This property was evident in Fig.~\ref{fig:narrow_oiii}: the peak emission of \oiiin\ for LBQS0109 is visible  E  of the AGN location (at 0,0 position that is identified with a red diamond) and  that of HB8903 is N and W of the AGN.
These narrow components are not detected in \hb\  probably because the S/N of the data is lower. 

\begin{figure}
   \centering
\includegraphics[width =0.24\textwidth]{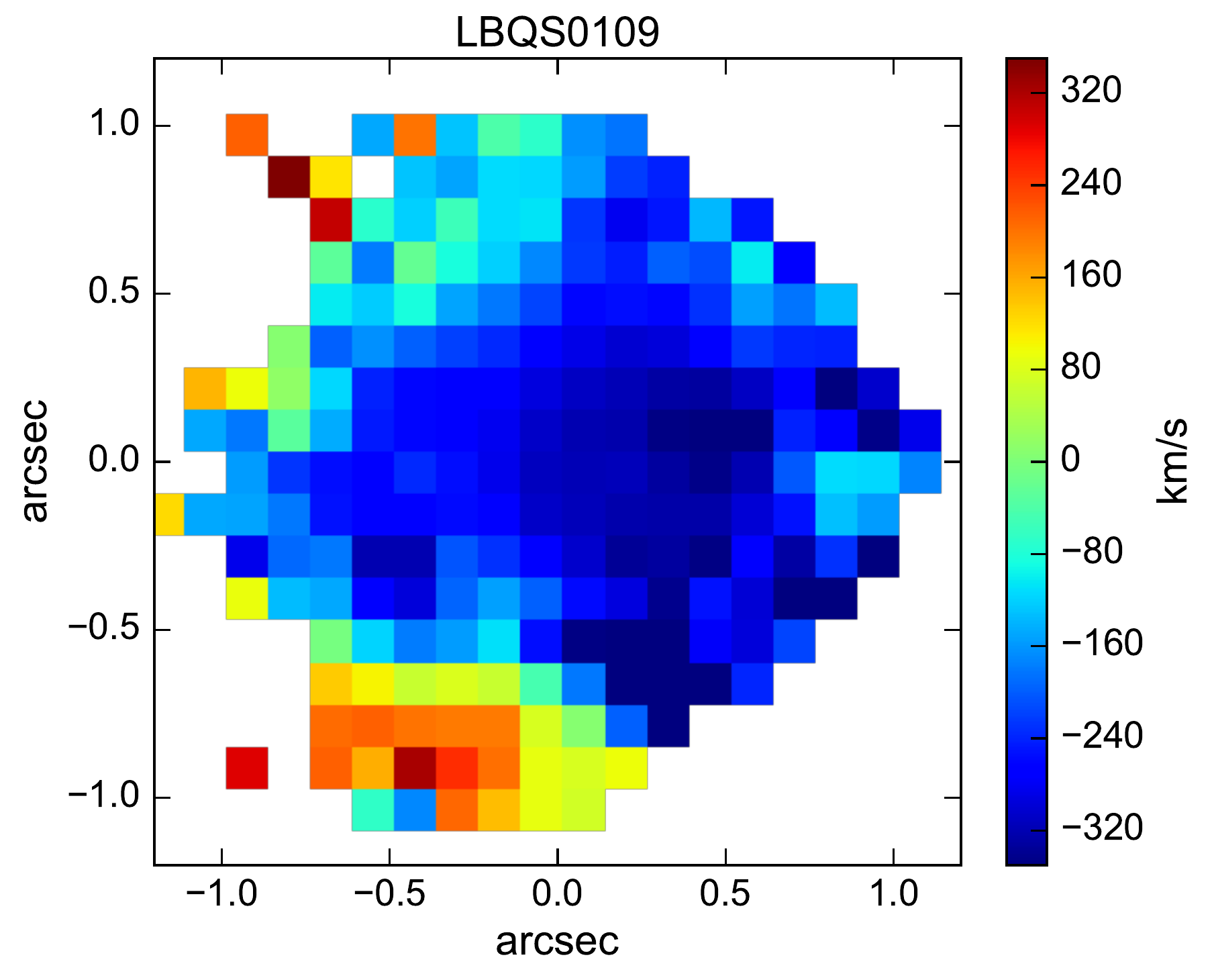}
\includegraphics[width =0.24\textwidth]{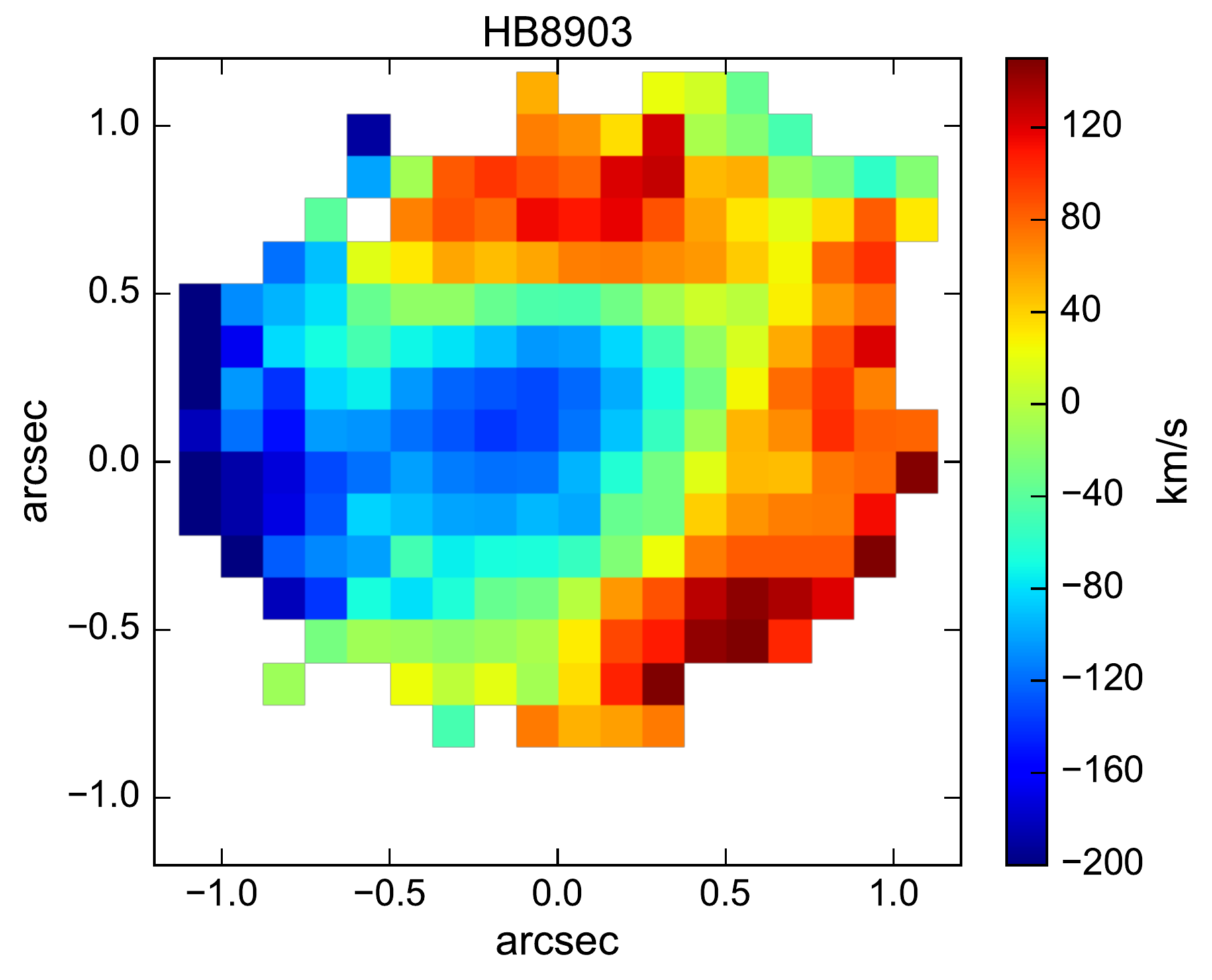}
\caption{  \oiii\ median velocity maps of LBQS0109 (left) and HB8903 (right). The velocity maps are obtained considering only the broad \oiii\ component, and they are therefore not influenced by the narrow \oiii. 
They are to be compared with the median velocity maps obtained for the overall \oiii\ profile shown in Fig.  \ref{fig:narrow_oiii}.}
 \label{fig:map_oiii_2}
   \end{figure}

\begin{figure}
   \centering
\includegraphics[width =0.4\textwidth]{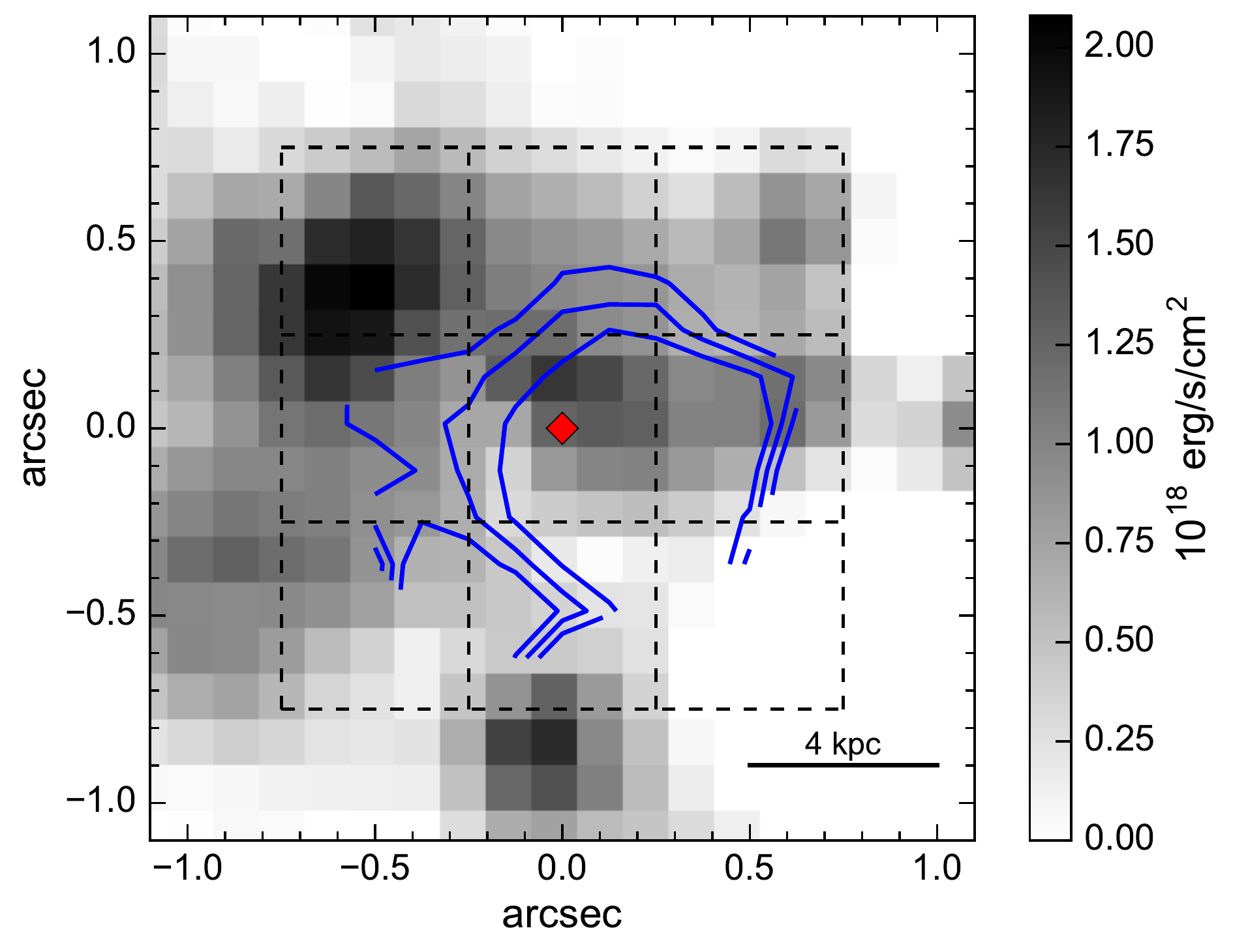}\\
\includegraphics[width =0.4\textwidth]{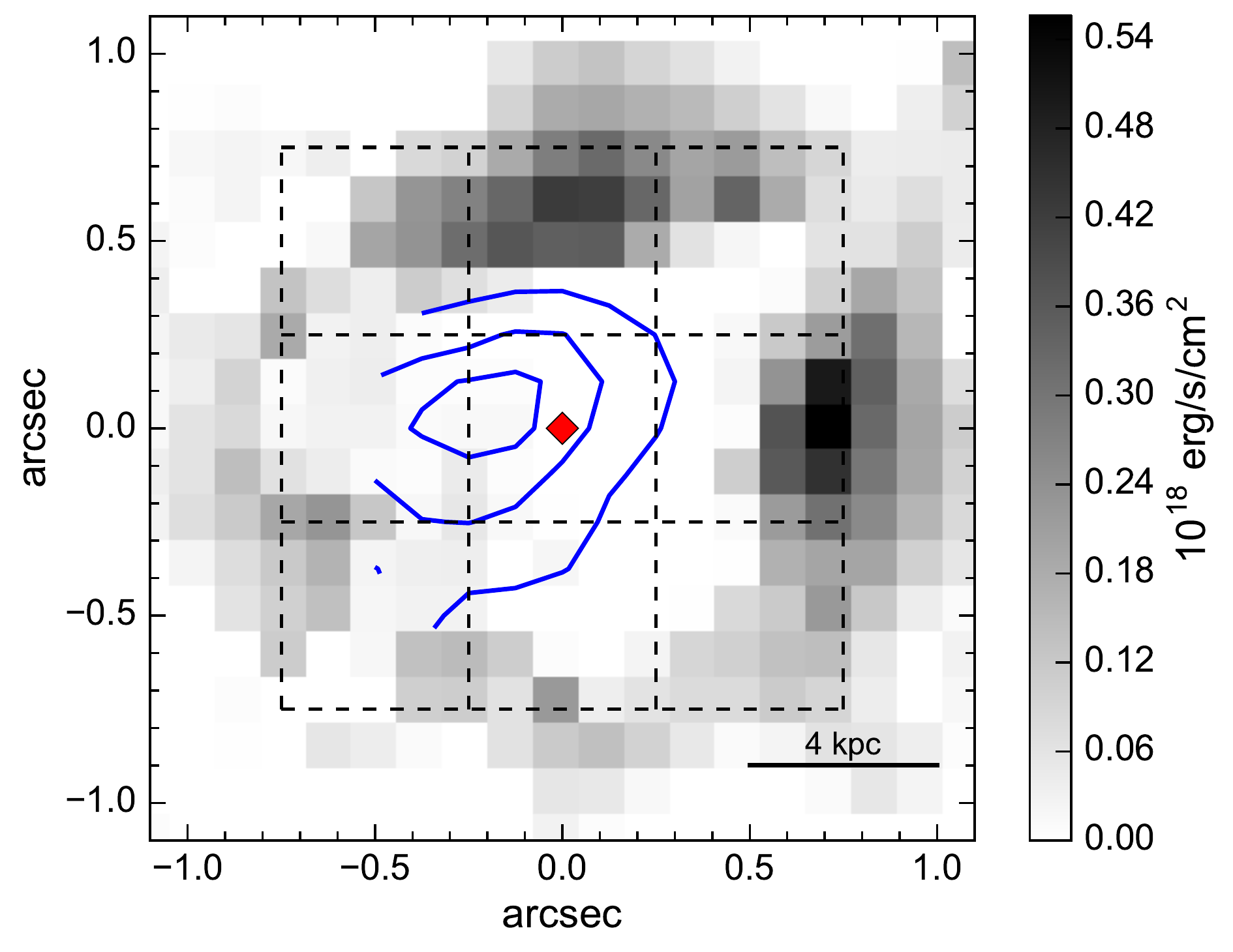}
\caption{ Flux maps of the \oiiin\ component for LBQS0109 (top) and for HB8903 (bottom)  obtained from the fit described in the text.
The AGN position is identified with a red diamond.  The blue  contours trace the \oiii\ blue-shifted emission cone shown in Fig.~\ref{fig:map_oiii_2}. The contours represent the velocity -300, -275 and -250 km/s for LBQS0109 , and -125,-100 and -50 km/s for HB8903. The \oiiin\ emissions are not symmetrically distributed and are anti-correlated with the outflow region. 
 Black squares indicate the nine regions where we extracted the corresponding spectra shown in the right panels of Fig.~\ref{fig:narrow_oiii}.
}
 \label{fig:map_oiii}
   \end{figure}

 The kinematics of the broad \oiii\ components obtained after subtracting  \oiiin\ are shown in Fig.~\ref{fig:map_oiii_2}. They are similar to those shown in  \pap\ and in the left panels of Fig.~1, indicating that they are not significantly affected by the narrow components. 

The comparison of the \oiiin\ flux maps with the velocity maps of the broad \oiii, whose negative velocity are shown in blue contours in Fig. \ref{fig:map_oiii}, reveals that the emission of the two narrow components is not spatially correlated with the ionised outflows.  The surface brightness of  \oiiin\ seems to be attenuated in the S-W region of LBQS0109 and in the S-E region of HB8903: both regions  are characterised by high blue-shifted velocity associated with the ionised outflow approaching along the line of sight. 

By applying the spectroastrometric method to the \oiii\ emission line, we found in Paper I that  the photocentre position extracted from the blue side of the line profile  is displaced with respect to the QSO position (Fig. 4 of \pap). This offset is the signature of  outflowing gas moving away from the nuclear region. In several cases,  including the two QSO presented here, we also observed an offset on the red side of the line profile, indicating that  ionised gas at red-shifted velocities is also emitted on galactic scales, far from the QSO position. 
 Since \oiiin\ emission detected in these two targets peaks  in the same region (N-E for LBQSO0109 and W for HB8903) as the position of the red photocentres,   it is very likely that the \oiiin\ is the origin of the displacement observed on the red side. 
A narrow component may then be present  in the other sources of our sample, but it is too weak to be clearly deblended {\rm either} from the stronger nuclear \oiii\ component {\rm or from the broad \oiii\  outflow component, which may contaminate the red side profile.}

The \oiiin\ spatial extension well over 4 kpc excludes the possibility that this line is emitted by NLR, which usually extends from a few hundred pc to a few kpc. \citep[see e.g.][]{Netzer:2004, Husemann:2014, Young:2014}.
Both the line widths (FWHM$\sim 480$ km/s)  and the line strengths of the \oiiin\ compared to the main broad \oiii\ component  suggest that these narrow components  might be associated with star-forming regions in the galactic disks.

\subsection{K-band spectral fitting and the narrow \ha\ component}\label{sec:fitting}

\begin{figure*}
   \centering
\includegraphics[width =0.49\textwidth]{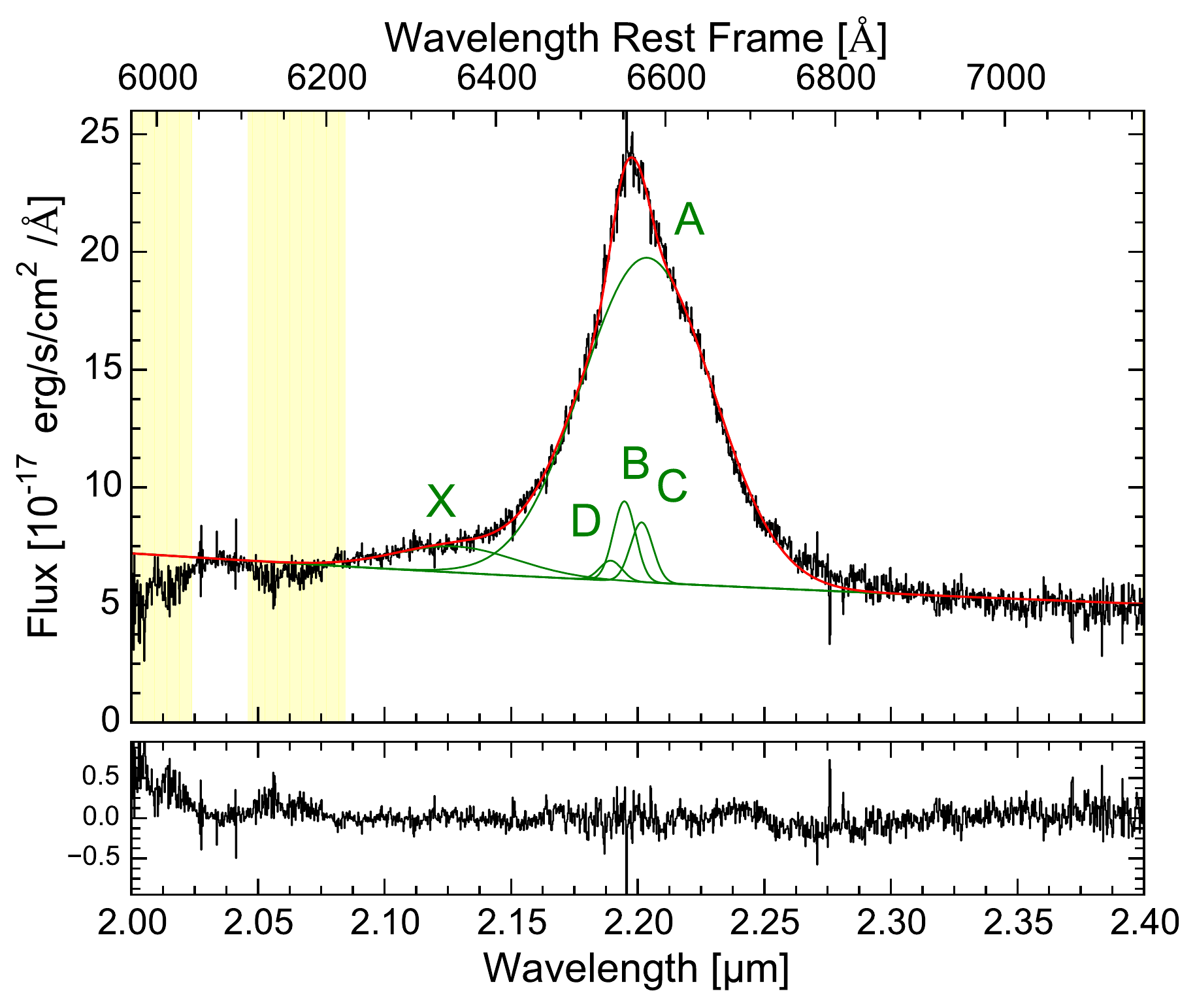}
\includegraphics[width =0.49\textwidth]{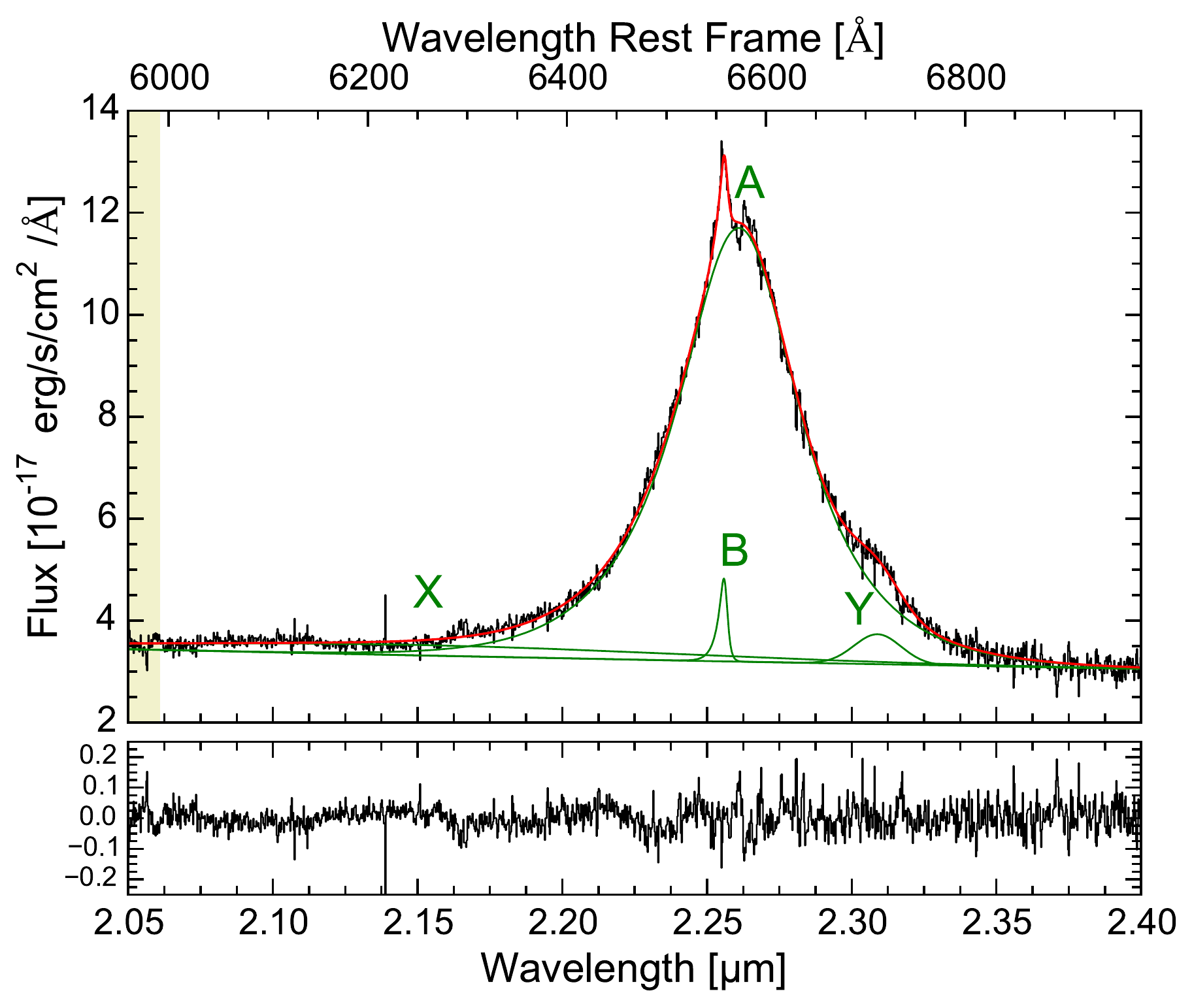}
\caption{ K-band SINFONI spectra of LBQS0109 (left panel) and  HB8903  (right panel) in the spectral region of \ha\, integrated in a region of 0.25\arcsec$\times0.25$\arcsec\ (2 $\times$ 2 pixel) around the QSO. \emph {Upper panels}: the observed spectra are shown with a black line. The different components in the fit for each line (\ha, \nii,\  and continuum) are shown in green, and the red line shows the total fit. The shaded yellow regions indicate the zone affected by strong sky line residuals, which are excluded from the fit. \emph {Bottom panels}: fit residuals, obtained as a difference between observed and model spectra.}
 \label{fig:spectra}
   \end{figure*}

Strong \ha\ emission is detected in both  sources. The spectra extracted from a nuclear region ($0.25\arcsec\times0.25\arcsec$) show a very broad \ha\ profile (Fig. \ref{fig:spectra}).
From comparing the \ha\ profile with the \oiii\ and \hb\ profiles analysed in \pap, we expect a broad (FWHM $>$ 4000 km/s) component originating from the BLR and a narrower  component (500 km/s $<$ FWHM $<$ 2000 km/s) overlapping with the \oiii\ outflow.
To set up the spaxel-by-spaxel fit for both targets, we initially considered the spectrum extracted from the nuclear aperture.
Continuum emission was fitted with a single power law, while the \ha\ line was fitted with broken power-law profiles for the BLR component and for the narrower NLR component associated with the outflow.  In Fig.~\ref{fig:spectra} the BLR and NLR components are labelled A and B, respectively.
The [NII] doublet associated with component B 
was fitted with two Gaussian components (labelled C and D), and their velocities and widths were fixed to those  of component B. Since the two \nii\ emission lines ($\lambda\lambda 6584, 6548$\,\AA) originate from the same upper level, their intensity ratio was fixed at the value of 3, given by the ratio of the Einstein coefficients. 
For HB8903, the fitting procedure did not require the addition of the [NII] components to reproduce the \ha\ profile.
In addition to the previously mentioned components (A, B, C, and D for LBQS0109 and A and B for HB8903), we added one Gaussian component for  LBQS0109  (component X in the left panel of Fig. \ref{fig:spectra}) and two Gaussian components for HB8903 (X and Y in the right panel of Fig. \ref{fig:spectra}). The Y component  may be associated either with the broad HeI line at $\lambda = 6679.99 \AA$ or the doublet [S{\sc ii}]$\lambda\lambda6717, 6731\AA$. The wavelength of  component X in both sources cannot be associated with any known emission line, but it was added to reduce the $\chi^2$ of our fit.
The fit residuals shown in Fig.~\ref{fig:spectra} indicate that the adopted components  describe the nuclear spectrum of each target well.

\begin{figure}
   \centering
\includegraphics[width =0.45\textwidth]{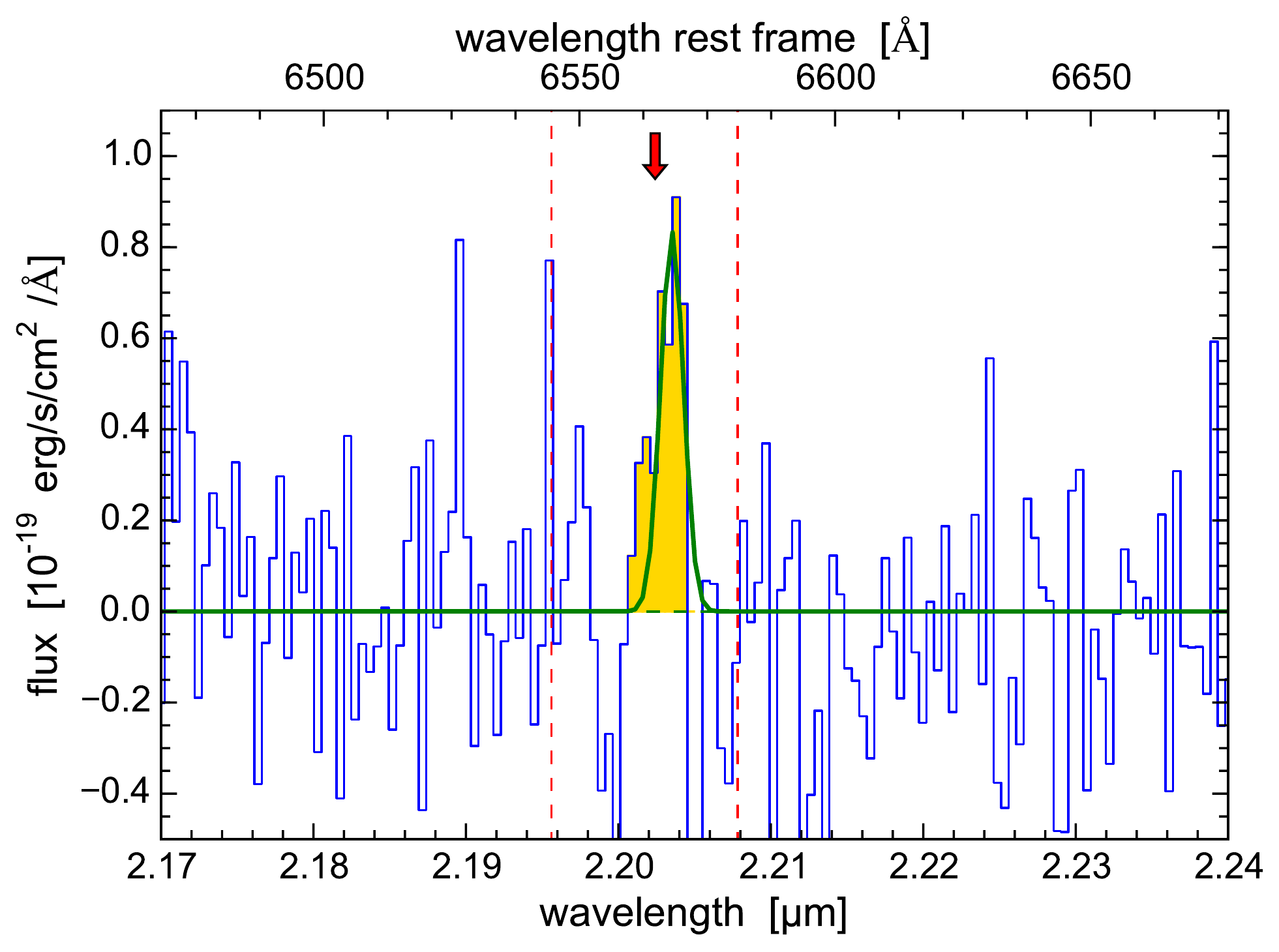}\\
\includegraphics[width =0.45\textwidth]{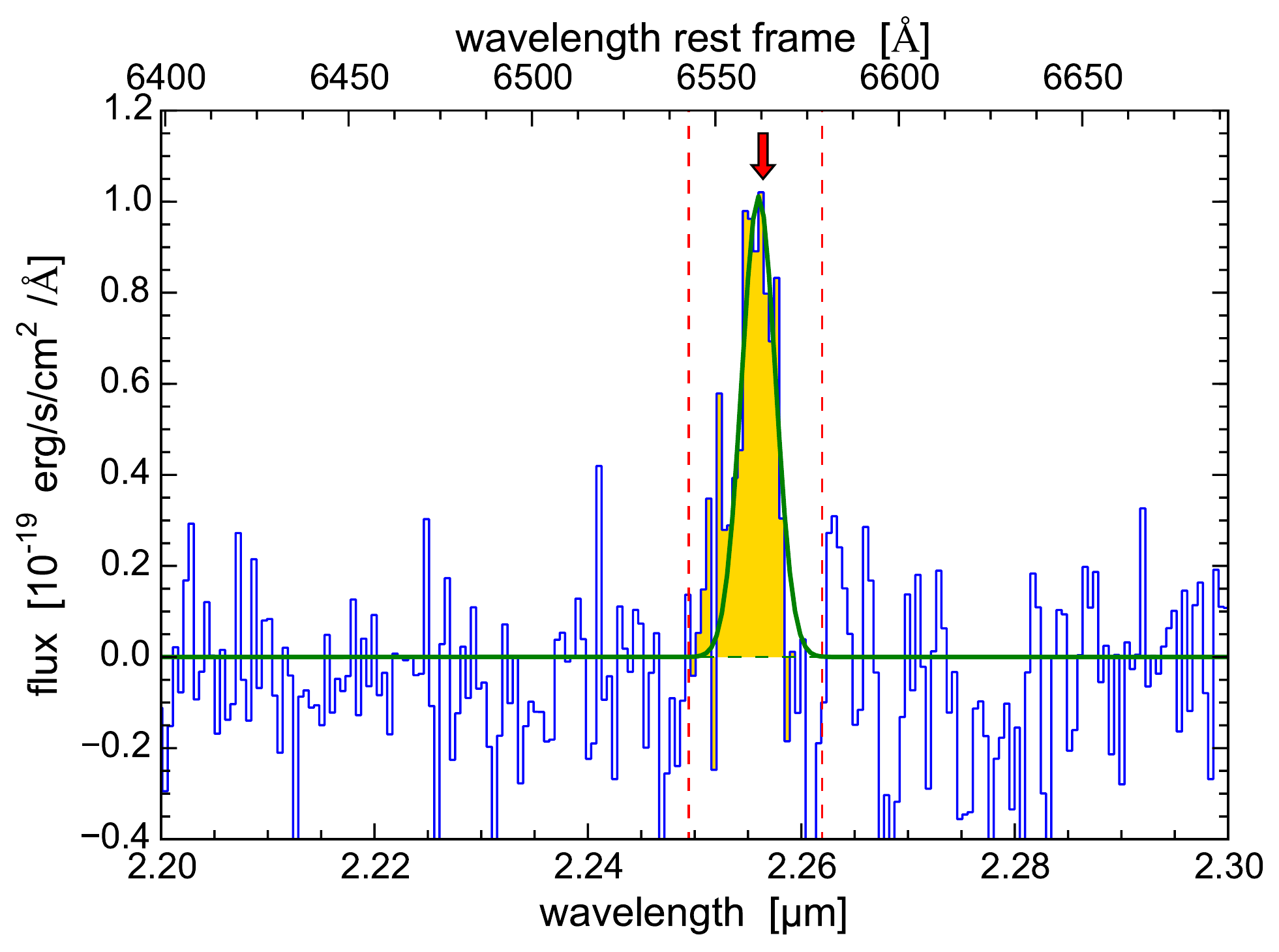}
\caption{ Narrow \ha\ emission from the residual K-band fitting,   extracted from  a ring-shaped region ( 0.5\arcsec$< r<$ 0.8\arcsec) of  LBQS0109 (top) and HB8903 (bottom). The red arrows indicate the expected position at the redshift of the \oiiin\ component. The dashed red line indicates the expected positions for doublet \nii: the lines are not detected, confirming that star formation is the excitation mechanism for the narrow \ha\ line (see text).}
 \label{fig:ha_narrow}
   \end{figure}

We performed the spectral fitting on individual spaxels by fixing the  kinematics of the broad components and by allowing the flux (i.e. the amplitude) to vary. Figure \ref{fig:ha_narrow}  shows 
 the residual spectra of the pixel-by-pixel fitting extracted from  a ring-shaped region ( 0.5\arcsec$<$ r$ <$ 0.8\arcsec) of the field of view.
These residual spectra reveal a weak but significant narrow (FWHM $\sim$ 250 km/s for LBQS0109 and FWHM$\sim$500 km/s for HB8903; see Table \ref{tab:results}) \ha\ component.  The velocity  of this narrow component (hereafter \han) is consistent within the errors  with the red-shift of the narrow \oiiin\ detected in the H-band spectra (Sect. \ref{sec:oiii}): the red arrow in Fig. \ref{fig:ha_narrow} denotes the velocity of \oiiin.
The red dashed lines show the expected position of the  corresponding narrow \niin\ components associated with   \han. 
The doublet \niin\ is clearly not detected  in the residual spectra of either object, resulting in an upper limit of  $\log_{10}$(\niin/\han) $<$ -0.85 for LBQS0109 and of $\log_{10}$(\niin/\han) $<$ -1.32 for HB8903.
By studying the line ratios using the Baldwin, Phillips \& Terlevich (BPT) method \citep{Baldwin:1981}, we find that the narrow emission is consistent with gas excited by the star formation region \citep{Kauffmann:2003}. 
Moreover, the FWHM$\lesssim 500$ km/s of the \han\ component is consistent with the typical line widths observed in  star-forming galaxies at z$\sim$2.5 \citep{Forster-Schreiber:2009}.
  
\subsection{SED fitting}\label{sec:SED_fitting}

  \begin{figure*}
   \centering
\includegraphics[width =0.45\textwidth]{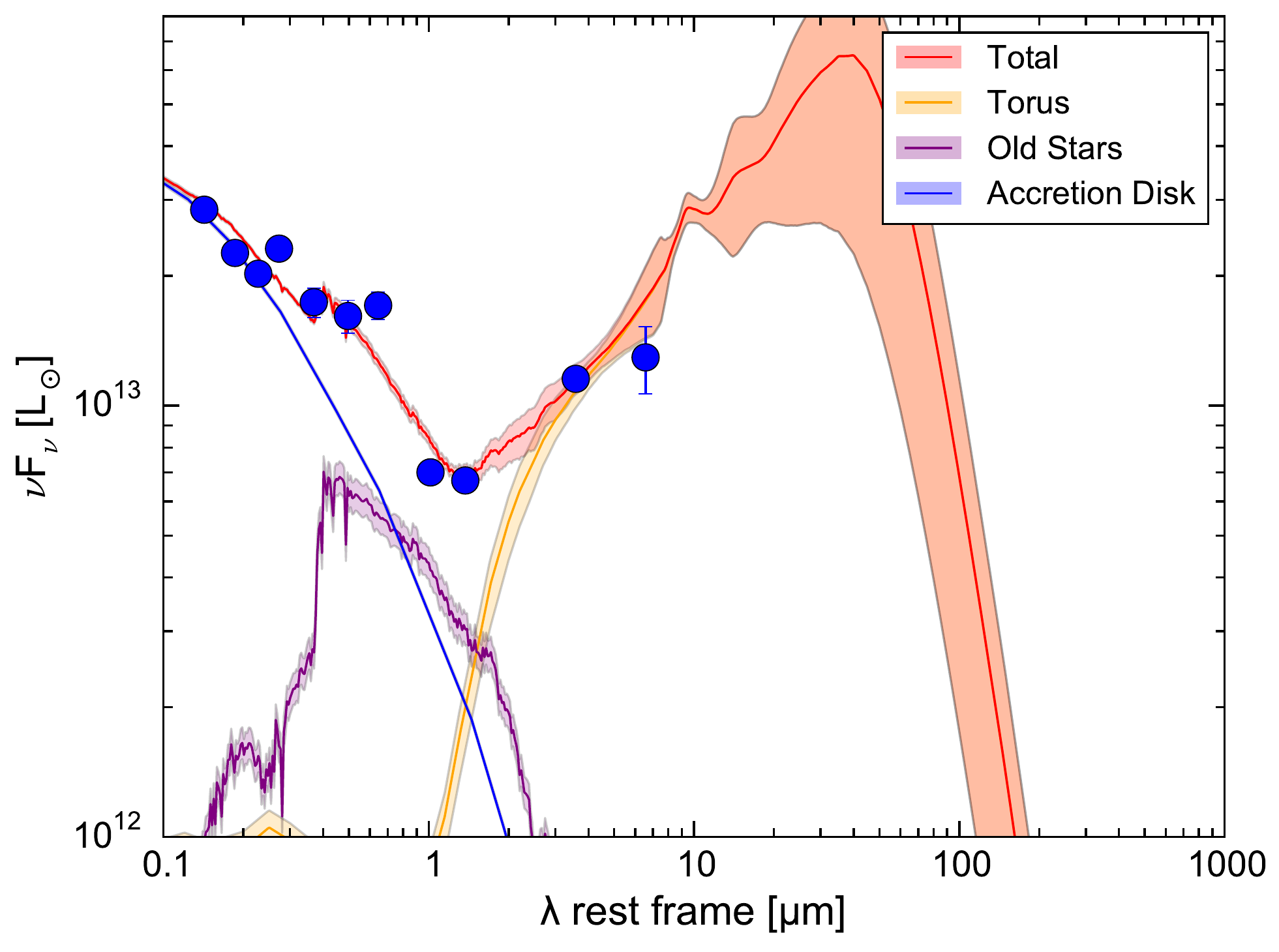} 
\includegraphics[width =0.45\textwidth]{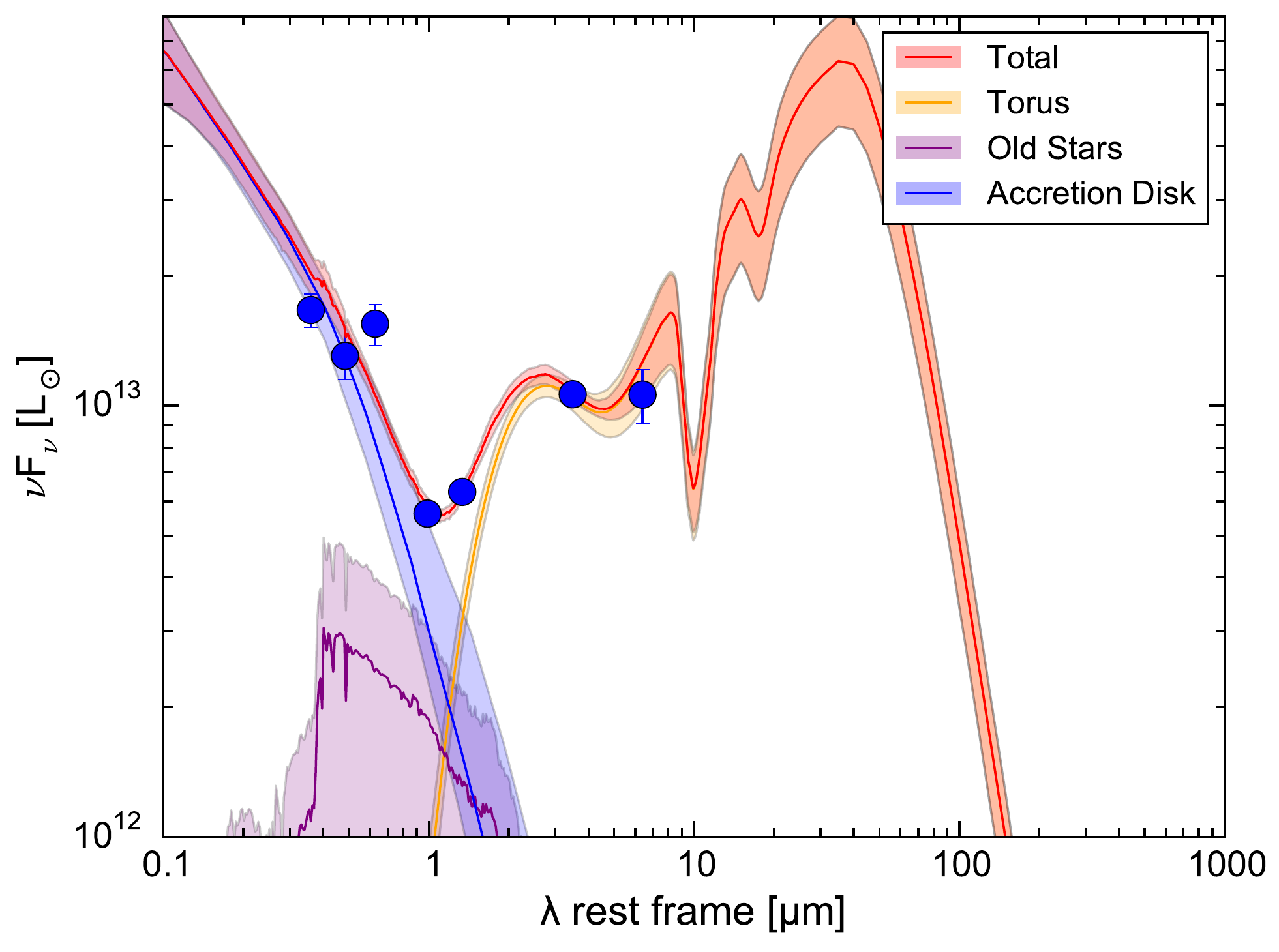}
\caption{SED fitting of LBQS0109 (left) and  HB8903 (right). The blue points show the  QSO emission fluxes from SDSS, 2MASS, and WISE photometry. 
The total best-fit models are plotted in red, while accretion disk, old star, and torus components are plotted in blue, purple, and orange, respectively.
The filled areas denote the uncertainties estimated with a Monte
Carlo procedure, see \cite{Balmaverde:2016} for more details.}
 \label{fig:SED}
   \end{figure*}

Figure \ref{fig:SED} shows the rest-frame spectral energy distributions (SEDs) of the two QSOs obtained by combining Sloan Sky Digital Survey (SDSS; when available), Two Micron All Sky Survey (2MASS) and Wide-field Infrared Survey Explorer (WISE) photometry.
We did not include photometric points at rest-frame wavelength $<1215.67 \ \AA$ since such emission is affected by intergalactic medium absorption.
We took advantage of 2MASS data in the J (1.24 $\mu$m), H (1.66 $\mu$m) and K (2.16 $\mu$m) bands. The photometric points were extracted from All Sky Point Source Catalogue.
We used the data observed with the infrared telescope WISE in the W1 (3.4 $\mu$m), W2 (4.6 $\mu$m), W3 (12 $\mu$m), and W4 (22 $\mu$m) bands. The continuum fluxes for both targets in all four bands can be found in the WISE All-Sky Source Catalogue. 
For LBQS0109,  we extracted the continuum flux from four out of five SDSS colour bands (g, r, i, and z), selecting the parameter \emph{psfMag} that measured the total flux of the object by fitting a PSF model in the position of the target.

The SED fitting decomposition, shown as a red solid line in Fig. \ref{fig:SED}, was performed by using the SED fitting  code  by \cite{Balmaverde:2016}, who assumed that the best-fit SED is the combination of different components: accretion disk emission,  unabsorbed (old) stellar population, emission of dust heated by the AGN (black body at $\sim 1500$K and torus template), and by a starburst emission (starburst template). We refer to \cite{Balmaverde:2016} for a detailed description of the adopted templates and fitting procedure.
The lack of photometric points in the rest frame 8-1000 $\mu$m does not allow us to constrain the presence of a possible starburst component. However, we note that such a component is not required by the SED fitting performed with the current data. 

The colour-filled areas indicate the uncertainties on the best fit, which were estimated through a Monte Carlo procedure, see \cite{Balmaverde:2016} for more details.
Clearly, the lack of photometric points beyond 10 $\mu$m rest-frame does not allow us to  constrain the emission from the AGN torus in either source.
The luminosity at $1-3\ \mu$m can be explained with the emission from an unobscured population of old stars in the host galaxy \citep{Meidt:2012}. From the SED fitting, we derive a stellar mass (hereafter \mstar) of $1.5\times10^{12}$ \msun\ for LBQS0109 and of $0.6\times10^{12}$ \msun\ for HB8903, the latter with a large uncertainty. However, since the old star components contribute less than 25\% to the total luminosity in the range between 1 $\mu$m and 3 $\mu$m, we consider the inferred \mstar\ as upper limits.

\section{Discussion}
In the following we discuss the interpretations of the two narrow components, \oiiin\ and \han, detected in the two QSOs. The surface brightness of \ha\ is similar to that observed in previous works \citep{Cano-Diaz:2012, Cresci:2015} and suggests two possible scenarios.

\subsection{Negative-feedback scenario}

  \begin{figure}
   \centering
\includegraphics[width =0.43\textwidth]{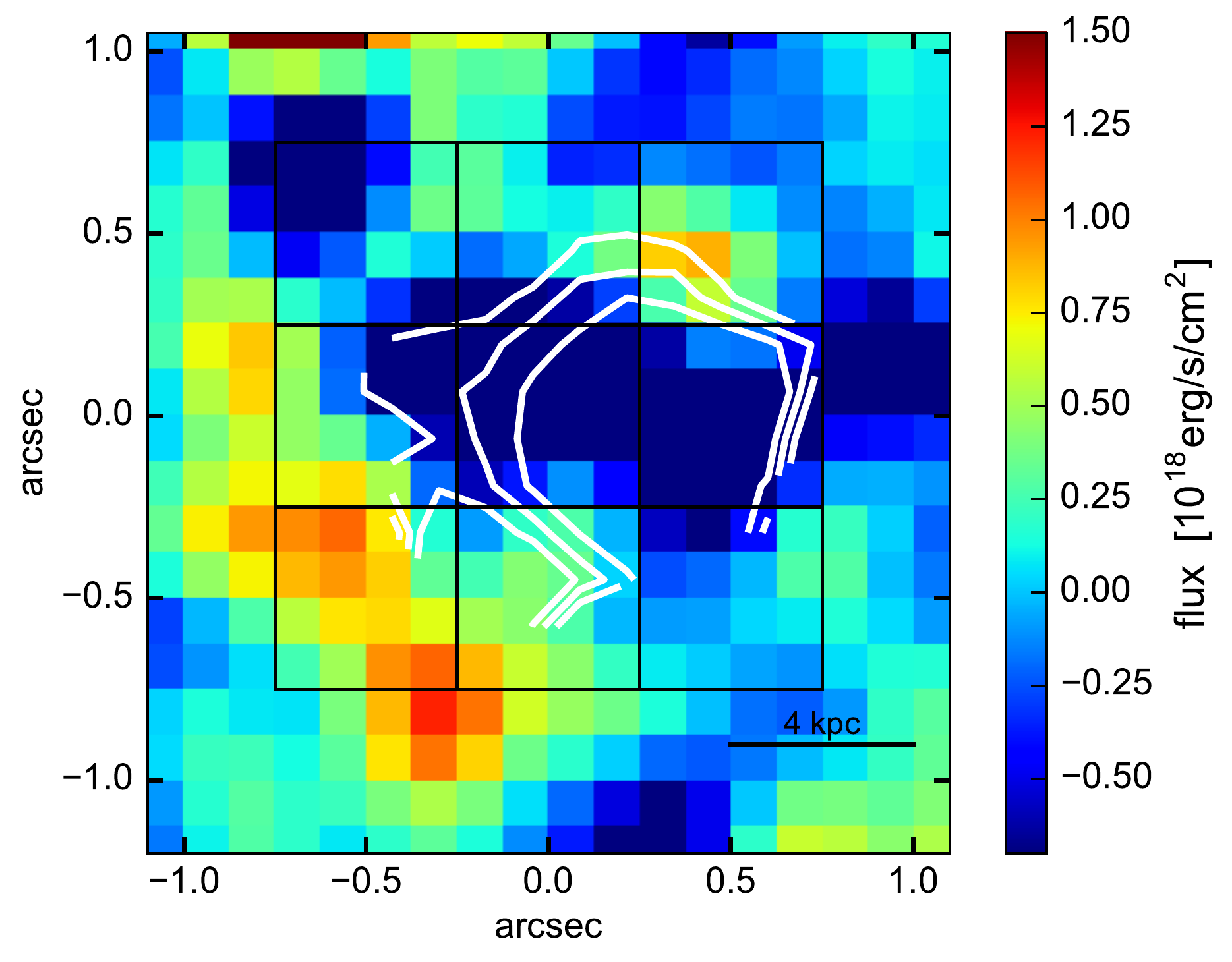} \\
\includegraphics[width =0.43\textwidth]{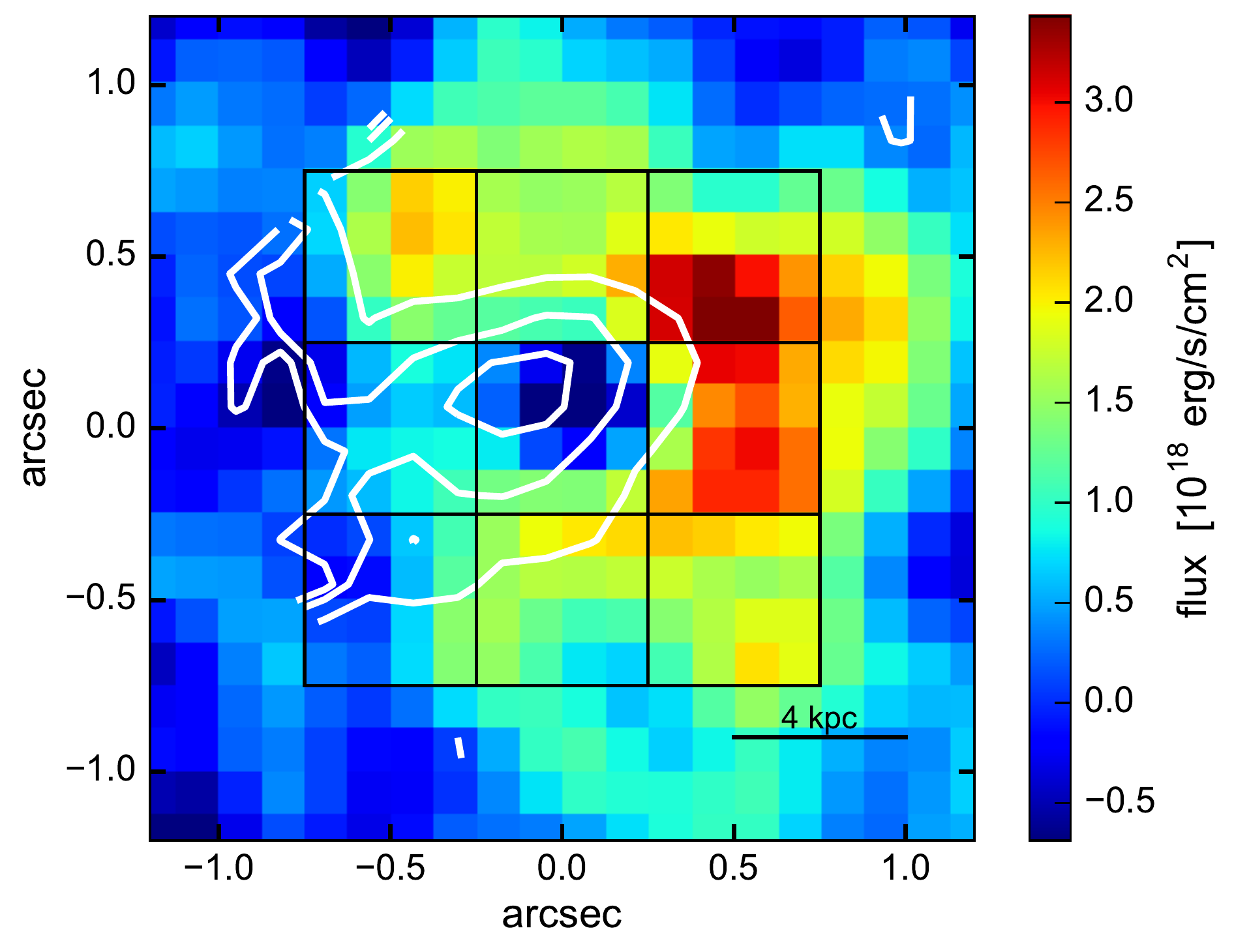}
\caption{ Maps of the narrow component of \ha\  for LBQS0109 (top) and HB8903 (bottom); the white  contours trace the \oiii\ velocity blue-shift  shown in Fig.~\ref{fig:map_oiii_2}. The contours represent the velocity  -300, -275, and -250  km/s for LBQS0109 , and -150, -100, and -50 km/s for HB8903. Star formation traced by \han\, is heavily suppressed in the S-W and S-E for region for LBQS0109 and HB8903, respectively, where the  outflow traced by \oiii\ is fastest.  Black squares indicate the nine regions where we extracted the corresponding H-band spectra shown in the right panels of Fig.~\ref{fig:narrow_oiii}.}
 \label{fig:ha_map}
   \end{figure}

Figure \ref{fig:ha_map} shows the flux maps of the \han\ component obtained by collapsing the residual data cubes of LBQS0109 and for HB8903  over the wavelength bins where \han\ is detected. The \han\ emission is extended to $\sim$1\arcsec\ ($\sim$ 8 kpc) from the AGN in both sources. Moreover, the \han\ surface brightness is   asymmetrically distributed around the nucleus, similarly to \oiii. The \han\ emission in LBQS0109 is primarily distributed toward the S-E, while that of  HB8903 is concentrated in three different clumps located N-W and S-W of the AGN position. The \han\ surface brightness maps are clearly similar to those obtained for  \oiiin\ shown in Fig. \ref{fig:map_oiii}.

The white contours in Fig.~\ref{fig:ha_map} show the most blue-shifted \oiii\ projected velocities (-200, -180, and -150 km/s for LBQS0109 , and -150, -100, and -50 km/s for HB8903) and therefore correspond to the regions where the fastest \oiii\ outflows are located:  these appear spatially anti-correlated with the \han\ emission. We interpret these observations as evidence for negative feedback in action because they indicate that  star formation is quenched in the regions where AGN-driven outflows interact with the host galaxy.  
The present data do  not allow us to distinguish between the possibilities that in the regions encompassed by the fast outflows, star formation is absent due to missing molecular gas or because
molecular gas has been  heated or made more turbulent. Our planned ALMA observations of the CO(3-2) line at a resolution similar to the SINFONI observations will address this question.

We now try to assess whether these AGN-driven outflows significantly
quench the SFR of the \textit{\textup{whole}} galaxy, beyond the regions shown in Fig. \ref{fig:ha_map}.

To establish whether the total star formation in the QSO host galaxies is significantly quenched, the location of these sources could be studied in the SFR-\mstar\ plane and be compared with the main sequence (MS) of star formation at $z\sim 2$: a significant quenching of star formation in the hosts of the two quasars could be revealed by these being significantly below the MS. 

We estimated the SFR in the quasar host galaxies from the \han\ emission, assuming a Chabrier initial mass function \citep{Chabrier:2003}. It was not possible to correct for reddening because we were
unable to obtain a reliable measurement or a significant upper limit on the narrow H$\beta$ emission. Therefore our SFR estimates are very likely lower limits. 
By using the  \cite{Kennicutt:2012} relation Log(SFR/\sfr) = Log(L$_{\rm H\alpha}$/erg s$^{-1}$) - 41.27, we derived a total SFR from \han\ of SFR$\sim50$ \sfr\ for LBQS0109 and SFR$\sim90$ \sfr\ for HB8903,  which is consistent with the wide ranges of SFR observed  in star-forming and QSO host galaxies at similar red-shifts \citep[e.g][]{Lutz:2008, Cano-Diaz:2012, Cresci:2015, Netzer:2016}. 
The detected \ha\ emission traces the instantaneous SFR, while most works that have tried to estimate the MS for SFR-stellar mass relation at high red-shift used far-infrared (FIR; $\lambda = 8-1000 \ \mu$m) or a combination of FIR and ultraviolet (UV) observations and therefore provided the time-integrated SFR, which can be considerably higher than the SFR obtained from \ha.

The SED fitting presented in Sect.~\ref{sec:SED_fitting} does not allow us to reliably estimate stellar masses but only upper limits, which are  $1.5\times10^{12}$ \msun\  and $6.0\times10^{11}$ \msun\ for LBQS0109 and HB8903, respectively.

From combining the uncertainty on the SFR with the upper limits on the stellar masses, we conclude that it is not possible to constrain the location of the quasar host galaxy sources on the SFR-\mstar\ plane. We only note that an MS star-forming galaxy with \mstar$=5\times10^{11}$~\msun\ at $z\sim2$  has an SFR$_{\rm MS}$ of $\sim$220 \sfr\ \citep{Whitaker:2012}. Since this is a factor $\sim2-4$  higher than the value we inferred from \ha, the two host galaxies are probably below the MS and, hence, may be quenched. However, the lack of FIR observations do not allow us to confirm this point.

Considering the uncertainties on SFR estimates,  the upper limits on  stellar masses, and  the intrinsic scatter of the MS of star formation, we conclude that it is not possible to establish  whether the AGN-driven outflows significantly quench star formation in the \textit{\textup{whole}} galaxy.

\subsection{Positive-feedback interpretation}

Our observations vaguely resemble the observations at a $z$=1.5 source (XID2028) by \cite{Cresci:2015}, where the ionised outflow was found to anti-correlate with the \ha\ emission and the rest-frame U-band surface brightness, which map the current star formation in the host galaxy. The two emissions lie in both edges of the outflow cone. These observations were interpreted in terms of both negative and positive feedback, where the  fast wind sweeps up the interstellar gas of the host along the core of the outflow (i.e. negative feedback) and the ionised outflow
simultaneously triggers star formation by compressing the gas clouds (i.e positive feedback).

In our sources and in that of \cite{Cano-Diaz:2012},  star formation is distributed along both edges of the outflow cone (Fig. \ref{fig:ha_map}) as shown in the schematic view in Fig. 5 of \cite{Cresci:2015}. Therefore, it is also possible to interpret our results in terms of positive feedback, which has been invoked by recent models  to explain the correlation between AGN and star formation activity in the host galaxy \citep{Silverman:2009, Imanishi:2011, Mullaney:2012, Zinn:2013, Zubovas:2013}.  
This scenario may be confirmed by the fact that \ha\ emission is mainly produced by  massive stars with lifetimes of $<10$ Myr. This time is as long as typical AGN activity, suggesting that the star formation activity is connected with AGN-driven outflows.  However, we cannot exclude that the current SFR in the host galaxies is a residual of SF activity occurring before the negative feedback phase. SFRs in the host galaxies of bright QSOs can be as high as 1000 \sfr, hence an SFR$<100$ \sfr\  might imply a significant reduction compared to the past. In this case, the existing star formation regions would not be induced by positive feedback, but would be regions of star formation that is not perturbed by AGN-driven outflows.

\section{Conclusions}

 The two QSOs LBQS0109 and HB8903, part of a sample of six presented in a previous paper \citep{Carniani:2015}, clearly show evidence of \oiii\ emission from the host galaxy in their SINFONI H-band spectra.  The \oiii\ emission is characterised by a broad blue-shifted component, tracing a powerful outflow (\pap), and by a narrow red-shifted component (FWHM $<$ 500 km/s) that is most likely associated with star-forming regions in the host galaxies.
Our new SINFONI K-band observations confirm this hypothesis. After subtracting pixel-by-pixel all broad (FWHM $>500$ km/s) \ha\ and [NII] components from K-band spectra, we detected faint narrow  (FWHM $\sim 250-500$ km/s) \han\ emission  at the same redshift of  \oiiin\ that is co-spatially distributed with this line in both  QSOs.

The emitted \han\ flux  is not symmetrically distributed around the location of the QSOs, but is  extended towards the S-E and W for LBQS0109 and HB8903, respectively.  
Line widths and  surface brightness distributions of these components both suggest that they are powered by SF in the host galaxies. 
The inferred upper limit on   Log(\nii/\ha) (-0.85 for LBQS0109 and -1.32 for HB8903) is  consistent with the typical value observed in star-forming regions, according to the standard BPT diagrams.
For LBQS0109, we cannot exclude the possibility that the \oiiin\ and \han\ emissions  are associated with AGN emission if we assume a metal-poor AGN at z$\sim$2.5 \citep[panel 2-d and 4-d of Fig. 5 of][]{Kewley:2013}.

Summarising, with this work we doubled the sample of QSOs (from two to four) where the effects of AGN-driven feedback on star formation are clearly visible. We note that in all these cases the SFRs inferred from the narrow H$\alpha$ emission is significant ($\sim 50-90$ \sfr) and,  as a result of the unknown dust obscuration on the narrow \ha, they  are very likely below the total SFR in the host galaxies. 

Although the statistics is still limited, the combined observations reinforce the idea that  quasar outflows do affect only a (small) part of the host galaxy; therefore either AGN feedback does not completely quench star formation, or several AGN episodes are needed to accomplish this.
A larger sample with similar \lagn\ and $z$ will allow us to reach more reliable results. In particular, a comparison of the SFR between sources that  show evidence for fast ionised outflows and those that do not will resolve the negative-feedback puzzle. 

On the other hand, our results may be consistent with the scenario  proposed by \cite{Cresci:2015}, in which outflows remove gas along the direction of motion and, simultaneously, compress the gas at the edges of the outflow cone, triggering star formation. Thus AGN-driven outflows might both quench and induce  star formation in  host galaxies.

\begin{acknowledgements}
We acknowledge financial support from INAF and the Italian Ministry of University and Research under the contracts PRIN-INAF-2011 (``Black Hole growth and AGN feedback through cosmic time'') and PRIN MIUR 2010-2011 (``The dark Universe and the cosmic evolution of baryons''). MB acknowledges support from the FP7 Career Integration Grant ``eEASy''  (CIG 321913).
RS acknowledges support from the European Research Council under the European Union (FP/2007-2013)/ERC Grant  Agreement n. 306476.
EP acknowledges financial support from INAF under the contract PRIN-INAF-2012. C.C. gratefully 
acknowledges support from the Swiss National Science Foundation Professorship grant PP00P2\_138979/1 (ETH Zurich).
We thank the anonymous referee for comments and suggestions that improved the paper.
\end{acknowledgements}

\bibliographystyle{aa} 
\bibliography{bibliography_sinfoni2}

\appendix

\end{document}